\documentclass[aps, prl, twocolumn, showpacs, superscriptaddress, floatfix]{revtex4}

\usepackage{graphicx}% Include figure files
\usepackage{dcolumn}% Align table columns on decimal point
\usepackage{bm}% bold math
\usepackage{times}
\usepackage{amsmath}

\begin{document}

\title{Spin Liquid Ground State in the Frustrated $J_1$-$J_2$ Zigzag Chain System BaTb$_2$O$_4$}

\author{A.A. Aczel}
\altaffiliation{author to whom correspondences should be addressed: E-mail:[aczelaa@ornl.gov]}
\affiliation{Quantum Condensed Matter Division, Oak Ridge National Laboratory, Oak Ridge, TN 37831, USA}
\author{L. Li}
\affiliation{Department of Materials Science and Engineering, University of Tennessee, Knoxville, TN 37996, USA}
\author{V.O. Garlea}
\affiliation{Quantum Condensed Matter Division, Oak Ridge National Laboratory, Oak Ridge, TN 37831, USA}
\author{J.-Q. Yan}
\affiliation{Department of Materials Science and Engineering, University of Tennessee, Knoxville, TN 37996, USA}
\affiliation{Materials Science and Technology Division, Oak Ridge National Laboratory, Oak Ridge, TN 37831, USA}
\author{F. Weickert}
\affiliation{MPA-CMMS, Los Alamos National Laboratory, Los Alamos, NM 87545, USA}
\author{V.S. Zapf}
\affiliation{MPA-CMMS, Los Alamos National Laboratory, Los Alamos, NM 87545, USA}
\author{R. Movshovich}
\affiliation{MPA-CMMS, Los Alamos National Laboratory, Los Alamos, NM 87545, USA}
\author{M. Jaime}
\affiliation{MPA-CMMS, Los Alamos National Laboratory, Los Alamos, NM 87545, USA}
\author{P.J. Baker}
\affiliation{ISIS Facility, STFC Rutherford Appleton Laboratory, Harwell Oxford, Oxfordshire, OX11 0QX, UK }
\author{V. Keppens}
\affiliation{Department of Materials Science and Engineering, University of Tennessee, Knoxville, TN 37996, USA}
\author{D. Mandrus}
\affiliation{Department of Materials Science and Engineering, University of Tennessee, Knoxville, TN 37996, USA}
\affiliation{Materials Science and Technology Division, Oak Ridge National Laboratory, Oak Ridge, TN 37831, USA}

\date{\today}% It is always \today, today,
             % but any date may be explicitly specified

\begin{abstract}
We have investigated polycrystalline samples of the zigzag chain system BaTb$_2$O$_4$ with a combination of magnetic susceptibility, heat capacity, neutron powder diffraction, and muon spin relaxation measurements. Despite the onset of Tb$^{3+}$ short-range antiferromagnetic correlations at $|\theta_{CW}|$~$=$~18.5~K and a very large effective moment, our combined measurements indicate that BaTb$_2$O$_4$ remains paramagnetic down to 0.095~K. The magnetic properties of this material show striking similarities to the pyrochlore antiferromagnet Tb$_2$Ti$_2$O$_7$, and therefore we propose that BaTb$_2$O$_4$ is a new large moment spin liquid candidate.
\end{abstract}

%\keywords{}

\maketitle

\renewcommand{\topfraction}{0.85}
\renewcommand{\textfraction}{0.05}
\renewcommand{\floatpagefraction}{0.9}

%Maybe replace "classical spin liquid" with "large moment spin  %liquids"
%Also note that Tb2Ti2O7 is referred to as a quantum spin ice or %quantum spin liquid in some literature. Can my system be       %referred to as a quantum spin ice or quantum spin liquid?

Spin liquids are exotic ground states of frustrated magnets in which local moments are highly correlated but still fluctuate strongly down to zero temperature\cite{10_balents}. In principle, the fluctuations of a spin liquid can be quantum or classical in nature. Several types of spin liquids have been proposed theoretically, including Anderson's resonating valence bond state\cite{87_anderson}, spin ice\cite{01_bramwell, 09_gingras} and others characterized by either gapped or gapless low-energy excitations\cite{02_wen}. The experimental search for new spin liquid candidates is an ongoing area of interest since this state of matter remains largely unexplored in the laboratory. Some well-known examples of quantum spin liquid candidates include ZnCu$_3$(OH)$_6$Cl$_2$(herbertsmithite)\cite{07_helton, 12_han}, BaCu$_3$V$_2$O$_8$(OH)$_2$(vesignieite)\cite{09_okamoto}, and Ba$_3$NiSb$_2$O$_9$\cite{11_cheng}, while their classical counterparts are the pyrochlore magnets Ho$_2$Ti$_2$O$_7$ and Dy$_2$Ti$_2$O$_7$\cite{97_harris, 99_ramirez, 10_balents}. 
    
Magnetic systems characterized by frustration and low-dimensionality have proven to be useful starting points in the quest for uncovering additional spin liquids, but particular complications have severely limited the number of viable candidates. For example, although one-dimensional (1D) magnets and two-dimensional Heisenberg systems are not expected to order due to the Mermin-Wagner theorem\cite{66_mermin}, most real low-dimensional materials are governed by weak exchange interactions in the other spatial dimensions and therefore exhibit magnetic order at low temperatures. Furthermore, while magnetic frustration can drastically suppress leading terms in the Hamiltonian that would be responsible for magnetic order, there are instances in which the sub-leading terms can still drive the system to an ordered ground state\cite{09_mentre}. To overcome these obstacles and find new spin liquid candidates, it is important to perform detailed studies on a wide variety of frustrated, low-dimensional magnets. 

The family of materials AR$_2$O$_4$ (A~$=$~Ba, Sr; R~$=$~rare earth)\cite{05_karunadasa, 06_doi, 14_besara} satisfy the two criteria described above. Two crystallographically-inequivalent R sites independently form two different types of zigzag chains running along the $c$-axis, and therefore quasi-1D magnetic behavior may be expected. Bulk characterization studies have also shown that most members of the family have dominant antiferromagnetic (AFM) exchange interactions and relatively large frustration indices. Geometric frustration can arise in this structure type if the $J_2$ exchange interactions are AFM and the $J_1$ couplings are of comparable strength. 
 
\begin{figure*}
\centering
\scalebox{1.22}{\includegraphics{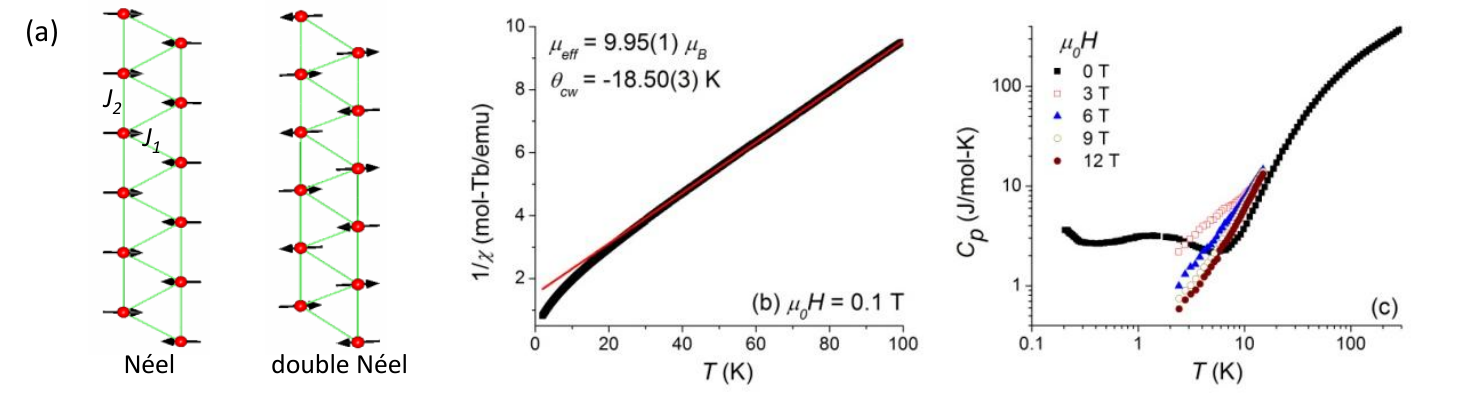}}
\caption{\label{characterization} (a) Schematics of the zigzag chain showing both the N\'{e}el and double N\'{e}el ground states that arise from the classical $J_1$-$J_2$ Ising model in the strong AFM $J_1$ and $J_2$ limits respectively. (b) Magnetic susceptibility of BaTb$_2$O$_4$, showing no evidence for long-range order down to 2~K. (c) $C_p$ data measured between 0 and 12~T are shown on a double logarithmic scale with no indication of magnetic order. The low temperature maximum at $T^*$~$=$~1.5~K in zero field is likely due to a low-lying crystal field level. }
\end{figure*}

The general trend for the AR$_2$O$_4$ family is that a small (large) ionic radius\cite{76_shannon} $r$ for the R atom shortens (lengthens) the $J_1$ bonds and induces N\'{e}el (double N\'{e}el) order on the chains. Schematics of these two magnetic ground states are shown in Fig.~\ref{characterization}(a). The N\'{e}el and double N\'{e}el long-range ordered states are consistent with predictions for the large AFM $J_1$ and $J_2$ limits of the classical Ising $J_1$-$J_2$ chain\cite{07_heidrich}, and have been realized in SrYb$_2$O$_4$($r$~$=$~0.87~\AA)\cite{12_quintero} and BaNd$_2$O$_4$($r$~$=$~0.98~\AA)\cite{14_aczel}  respectively. By tuning the ionic radii of the R atoms with different rare earths, the intermediate regime between these two limits has also been explored in detail. A variety of magnetic ground states have been observed, including coexisting long-range N\'{e}el and short-range double N\'{e}el order in SrEr$_2$O$_4$($r$~$=$~0.94~\AA)\cite{08_petrenko, 11_hayes}, coexisting short-range N\'{e}el and short-range double N\'{e}el order in SrHo$_2$O$_4$($r$~$=$~0.90~\AA)\cite{11_young, 13_young, 14_wen}, incommensurate magnetic order in SrTb$_2$O$_4$($r$~$=$~0.92~\AA)\cite{14_li}, short-range magnetic order in SrDy$_2$O$_4$($r$~$=$~0.91~\AA) at 0.05~K\cite{13_cheffings, 14_fennell}, and no magnetic ordering of any kind down to 0.065~K in SrTm$_2$O$_4$($r$~$=$~0.88~\AA)\cite{14_li_2}. It is also interesting to note that any long-range order observed in this family has been found to arise from only one rare earth site (except for SrYb$_2$O$_4$\cite{12_quintero}), and therefore the two types of zigzag chains are often characterized by different magnetic ground states. Since the $J_1$ and $J_2$ bond lengths are essentially equal for the two chain types, this behavior may be a consequence of the inequivalent, distorted oxygen octahedral local environments of the rare earths comprising each chain. 

In this Letter, we investigate polycrystalline BaTb$_2$O$_4$ ($r$~$=$~0.92~\AA) with a combination of magnetic susceptibility, heat capacity, neutron diffraction, and muon spin relaxation measurements. The magnetic species in this system is Tb$^{3+}$, which is a non-Kramers ion with a large angular momentum $J$~$=$~6. Despite the onset of antiferromagnetic correlations at the Curie-Weiss temperature $\theta_{CW}$~$=$~-18.5~K, there is no evidence for long-range magnetic ordering or spin freezing in any of the measurements down to 0.095~K. However, neutron diffraction reveals that incredibly short-ranged magnetic correlations exist between $J_2$ bonds. These findings provide strong evidence that BaTb$_2$O$_4$ is a new large moment spin liquid candidate. 

Polycrystalline BaTb$_2$O$_4$ samples were synthesized by a standard solid-state reaction method from high-purity starting materials of BaCO$_3$ and Tb$_4$O$_7$. First, Tb$_2$O$_3$ was obtained by reducing Tb$_4$O$_7$ in Ar (4\%~H$_2$). Next, a stoichiometric mixture of BaCO$_3$ and Tb$_2$O$_3$ (with 10\% excess BaCO$_3$) were ground, pressed into pellets, and then sintered in Ar (4\%~H$_2$) at 1150$^\circ$C for 8 hours. The final product was confirmed to be single phase by laboratory x-ray powder diffraction. 

The magnetic susceptibility of polycrystalline BaTb$_2$O$_4$ was measured in an applied field $\mu_0 H$~$=$~0.1~T under the zero-field-cooled condition using a Quantum Design Magnetic Properties Measurement System. The data are presented in Fig.~\ref{characterization}(b), plotted as 1/$\chi$, i.e. $\mu_0 H/M$, vs. $T$, and the results are in good agreement with previous work\cite{06_doi}. The high temperature data is well-described by a Curie-Weiss law, with a fit between 40 and 100~K yielding $\theta_{CW}$~$=$~-18.50(3)~K and an effective moment $\mu_{eff}$~$=$~9.95(1)~$\mu_B$. The effective moment is close to the expected value of 9.72~$\mu_B$ for Tb$^{3+}$. Despite the onset of AFM correlations around 20~K, there is no evidence for long-range magnetic order from the susceptibility measurements down to 2~K.

The specific heat ($C_p$) below 2~K was measured with a home-built probe based on the adiabatic heat-pulse technique in a He-3/He-4 dilution refrigerator from Oxford Instruments. Data at higher temperatures were taken in a Quantum Design Physical Properties Measurement System equipped with a 12~T superconducting magnet. Fig.~\ref{characterization}(c) shows $C_p$ data in selected applied fields. No evidence for magnetic order is found, while te observed field dependence is likely indicative of crystal field splitting that changes with $\mu_0 H$. 

The zero field $C_p$ data show a sharp upturn for $T$~$<$~0.3~K that can be attributed to nuclear Schottky contributions of Tb nuclei, as discussed for other Tb compounds\cite{73_catanese, 69_blote}. A broad maximum is also observed at $T^*$~$=$~1.5~K, which likely corresponds to a low-lying crystal field level. Integrating $C_p/T$ from 0.5-6~K over the broad peak(assuming a negligible lattice contribution in this range\cite{08_petrenko}) yields an entropy of just over $R$ln(3/2)/mol-Tb.  This finding could imply a doublet ground state for the Tb$^{3+}$ ions with a small energy gap to a low-lying singlet excited state. However, we note that $C_p$ measurements on the AR$_2$O$_4$ family provide limited quantitative information on the crystal field level schemes, due to the monoclinic site symmetries and the inequivalence of the local environments for the two different R sites. The build-up of short-range magnetic correlations with decreasing $T$ can also complicate analysis of the magnetic entropy extracted from $C_p$ data\cite{11_gaulin}.

\begin{figure*}
\centering
\scalebox{0.88}{\includegraphics{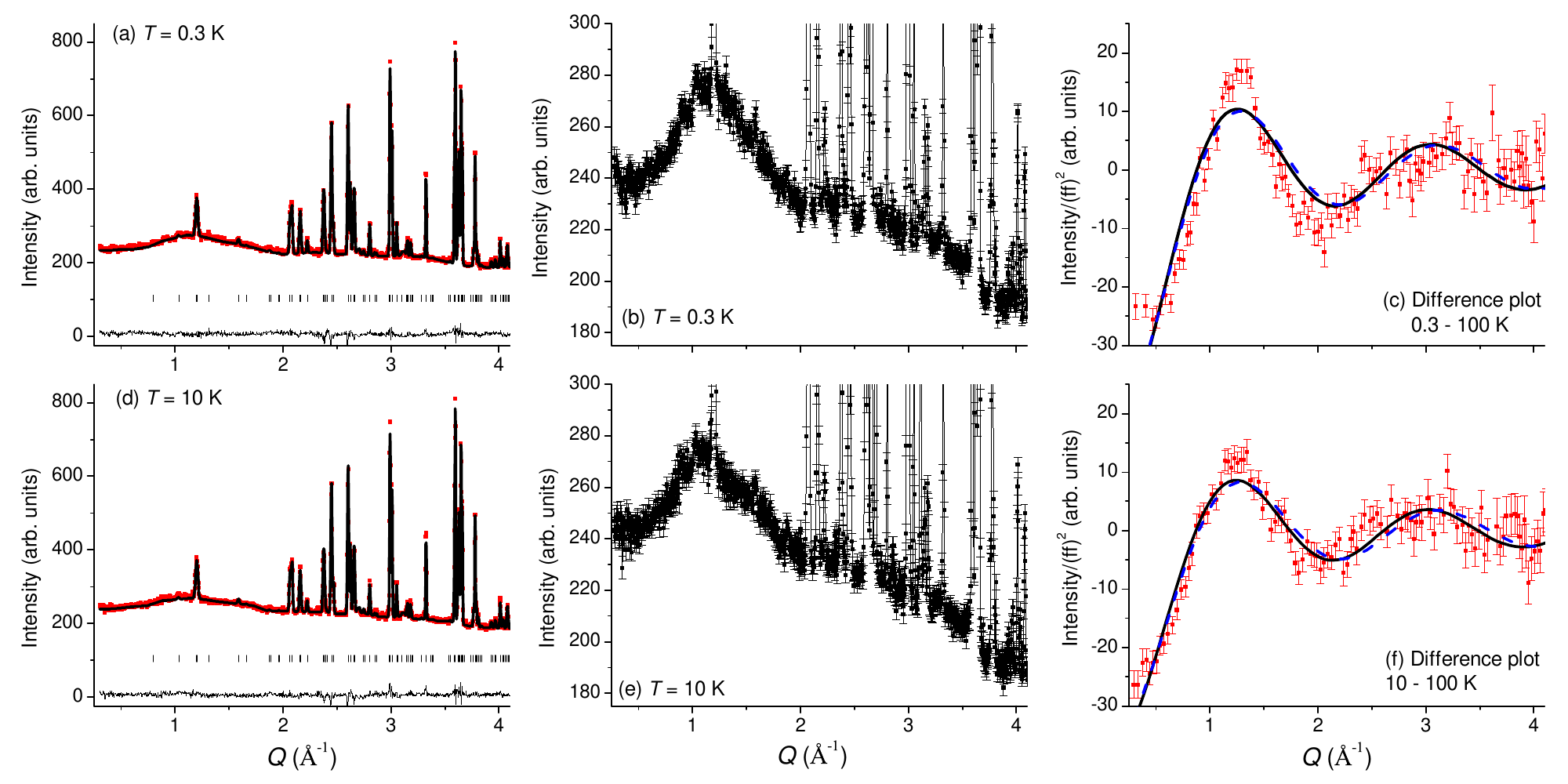}}
\caption{\label{neutron_diffraction} (a) HB-2A neutron diffraction data with $\lambda$~$=$~2.41~\AA~at 0.3~K for polycrystalline BaTb$_2$O$_4$. The solid curve is a fit generated from a structural Rietveld refinement using the space group {\it Pnam}. (b) An enlarged version of the data shown in panel (a), emphasizing the magnetic diffuse scattering. (c) A 0.3-100~K difference plot, with the intensity normalized by the Tb$^{3+}$ magnetic form factor squared. The dashed and solid curves are fits to models incorporating only Tb$^{3+}$ magnetic correlations between $J_1$ and $J_2$ bonds respectively. (d)-(f) Similar plots to those shown in panels (a)-(c), but with $T$~$=$~10~K.}
\end{figure*}

Neutron powder diffraction was performed with 5~g of polycrystalline BaTb$_2$O$_4$ between 0.3-100~K at Oak Ridge National Laboratory using the HB-2A powder diffractometer of the High Flux Isotope Reactor with a collimation of 12$'$-open-6$'$. 
Data with a neutron wavelength of 2.41~\AA~is depicted in Fig.~\ref{neutron_diffraction}(a) and (d) with $T$~$=$~0.3~K and 10~K respectively. Successful Rietveld refinements were performed using FullProf\cite{93_rodriguez} with the known room temperature space group {\it Pnam}\cite{06_doi}, indicating that there are no structural phase transitions down to 0.3~K. The lattice constants at 0.3~K refined as $a$~$=$~10.423(1)~\AA, $b$~$=$~12.178(1)~\AA, and $c$~$=$~3.497(1)~\AA. 

No evidence was found for long-range order in the diffraction data of BaTb$_2$O$_4$ . However, magnetic diffuse scattering was observed instead in both the 0.3 and 10~K datasets. This contribution is modeled as background in Fig.~\ref{neutron_diffraction}(a) and (d), and most clearly seen in Fig.~\ref{neutron_diffraction}(b) and (e). The $Q$-dependence of the diffuse scattering remains almost unchanged up to 10~K. These findings indicate that there are significant magnetic correlations that persist well above the onset of any possible long-range order. We note that the combined diffraction and susceptibility data rule out a well-isolated crystal field singlet ground state as found for Tm$_2$Ti$_2$O$_7$\cite{96_zinkin} and possibly applicable to SrTm$_2$O$_4$\cite{14_li_2}, since the signatures for such a scenario are a constant low-$T$ magnetic susceptibility and the absence of any elastic magnetic scattering, in contrast to observations.     

0.3-100~K and 10-100~K difference plots are shown in Fig.~\ref{neutron_diffraction}(c) and (f), with the data normalized by the Tb$^{3+}$ form factor squared. Similar oscillatory scattering patterns are clearly evident in both datasets, therefore no drastic change is found in the magnetic correlations through $T^*$. This observation confirms that the $C_p$ anomaly at $T^*$ is not associated with any form of magnetic ordering, but can likely be attributed to a low-lying crystal field level. The intensity of the difference plots is well-described by the function: 
\begin{equation}
I(Q)=\sum_{ij} A_{ij} \frac{sin(Qd_{ij})}{Qd_{ij}}
\end{equation}
This equation represents the expected polycrystalline response for a local magnetic structure with the spins at sites $i$ and $j$ correlated over distances of $d_{ij}$ only. Antiferromagnetic (ferromagnetic) correlations are inferred from $A_{ij}$~$<$~0 ($A_{ij}$~$>$~0). In BaTb$_2$O$_4$, the 0.3~K refinement discussed above yields a $J_1$ bond length of 3.50~\AA, while the two types of zigzag chains are found to have inequivalent $J_2$ bond lengths of 3.59~\AA~and 3.60~\AA. The best fits of the BaTb$_2$O$_4$ data, shown by the solid curves in Fig.~\ref{neutron_diffraction}(c) and (f), require only one term with $A$~$<$~0 and $d$~$=$~3.58(4)~\AA~at 0.3~K (3.61(3)~\AA~at 10~K). Fits with $d_{ij}$ fixed to 3.50~\AA~are also shown for comparison purposes by the dashed curves, but they do simulate the data as accurately. Therefore, the analysis described above is consistent with AFM correlations extending only between $J_2$ bonds, with $d$ representing an average bond length between the two types of chains. Furthermore, the incredibly short-range nature of the correlations is reminiscent of the magnetic behavior in the large moment spin liquid candidate Tb$_2$Ti$_2$O$_7$, where the Tb$^{3+}$ spins were found to be correlated over a single tetrahedron only\cite{99_gardner}.

The neutron diffraction results discussed above do not allow one to determine whether the spins are static or dynamic in the magnetic ground state. For example, diffuse elastic magnetic scattering was observed for all three pyrochlore systems Y$_2$Mo$_2$O$_7$\cite{99_gardner_2, 14_silverstein}, Tb$_2$Mo$_2$O$_7$\cite{92_gaulin}, and Tb$_2$Ti$_2$O$_7$\cite{99_gardner}, but the molybdates have been characterized as spin glasses while the titanate is best described as a spin liquid. Muon spin relaxation ($\mu$SR) measurements proved to be instrumental in unambiguously determining the nature of the ground states in these cases. More specifically, the $\mu$SR spectra for the two scenarios evolve quite differently in a longitudinal magnetic field configuration\cite{97_dereotier}. 

\begin{figure}
\centering
\scalebox{0.55}{\includegraphics{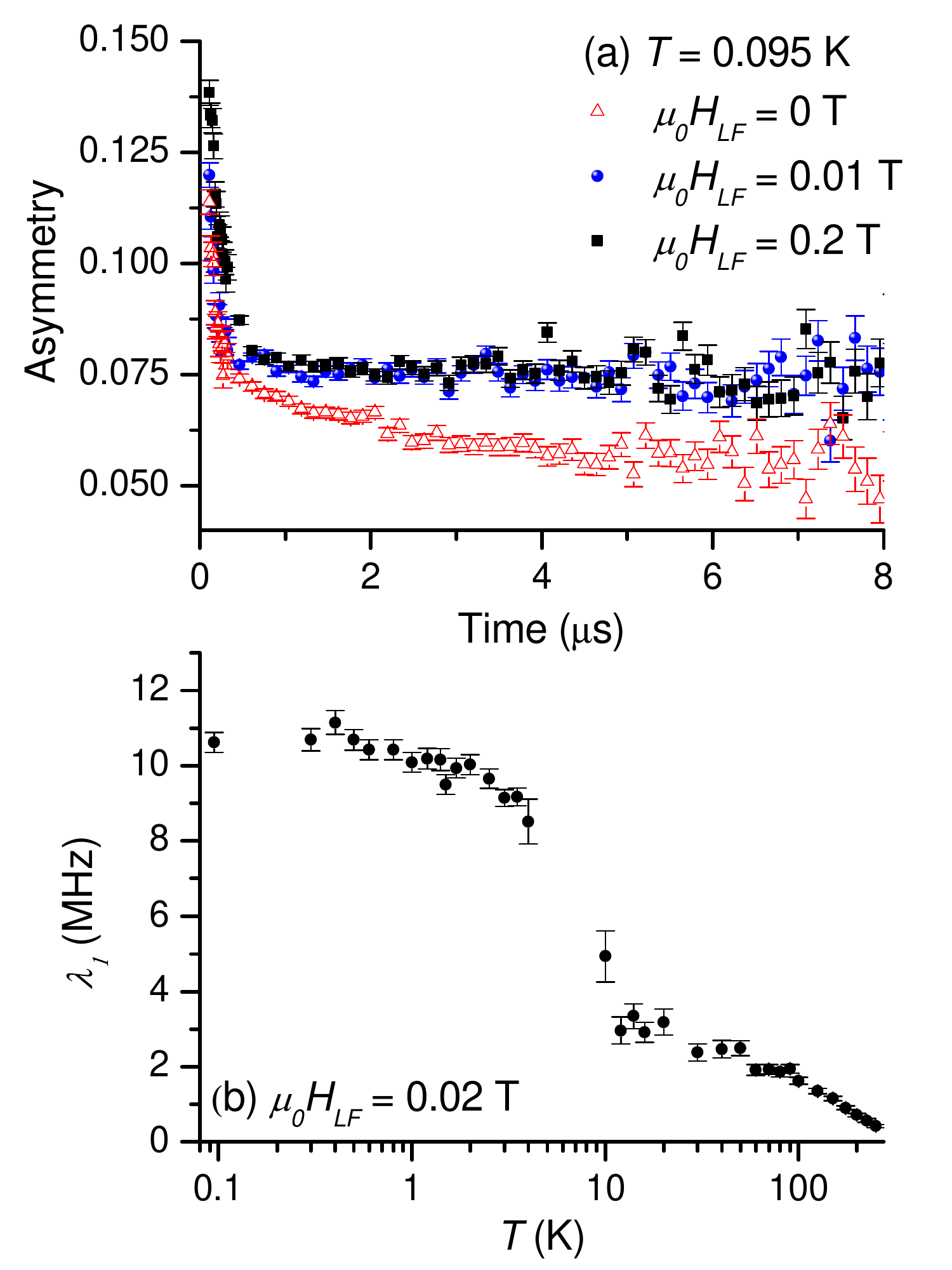}}
\caption{\label{muSR}(a) $\mu$SR spectra collected in selected LFs at 0.095~K. Very little decoupling of the zero field relaxation is observed up to 0.2~T, which suggests that its origin is a dynamic mechanism. (b) $T$-dependence of the relaxation rate $\lambda_1$, which saturates below 1~K. This $T$-dependence is consistent with dynamic behavior of the spins down to the lowest temperatures investigated.}
\end{figure}

Therefore, to better understand the magnetic ground state of BaTb$_2$O$_4$, $\mu$SR was performed at the EMU (10-300~K) and MuSR (0.095-4~K) spectrometers in longitudinal field (LF) geometry at the ISIS Pulsed Neutron and Muon Source. Zero field data were collected at 0.095, 0.5, 2, and 4 K, and no features characteristic of long-range order or spin freezing were observed. Data were also collected in several different LFs at 0.095~K. Some selected spectra are shown in Fig.~\ref{muSR}(a), and they fit well to the expression:
\begin{equation}
A(t)=A_1 e^{-\lambda_1 t} + A_{B} e^{-\lambda_{B} t} 
\end{equation}
where the two terms represent muons that stop in the sample and the sample holder respectively. The single sample component suggests that both types of chains host the same magnetic ground state, in contrast to the picture that has emerged for most other AR$_2$O$_4$ systems. This is consistent with the diffraction response that can be effectively modeled with only one local magnetic structure. 

One can estimate the local field at the muon site in the sample, assuming it is static, from the zero field $\mu$SR data by using the approximation: $B_{loc}$~$=$~$\lambda_1$/$\gamma_{\mu}$, where $\gamma_{\mu}$~$=$~135.5~$\times$~2$\pi$~MHz/T is the muon gyromagnetic ratio. In the present case, this analysis yields $B_{loc}$~$=$~0.0125~T for the lowest temperatures studied. Typically, if the observed muon spin relaxation in zero field is from a static mechanism, a LF up to one order of magnitude larger than the local field should be enough to completely decouple the relaxing signal. Fig.~\ref{muSR}(a) shows that the relaxation from the sample holder contribution is easily decoupled and therefore $\lambda_{B}$~$=$~0 for the LF spectra presented in this work. However, inspection of Fig.~\ref{muSR}(a) shows that the opposite is true for the relaxation associated with the sample contribution. In fact, the spectrum with $\mu_0 H_{LF}$~$=$~0.2~T shows only modest changes from the zero field case. This behavior is consistent with the muon spin relaxation arising from a dynamic origin and certainly not characteristic of a spin glass ground state.  

%It is worth pointing out that ZF data at different T's were    %measured - 0.09, 0.25, 0.5, 2, and 3.7 K - and no change was   %observed in the slow-relaxing part of the spectra.

$\mu$SR spectra were also collected with $\mu_0 H_{LF}$~$=$~0.02~T at a variety of different temperatures to allow for an accurate measurement of $\lambda_1$~$=$~$1/T_1$. Fig.~\ref{muSR}(b) depicts the $T$-dependence of $1/T_1$. No sharp peak is observed as would be expected in the case of spin freezing\cite{96_dunsiger}. Instead, $1/T_1$ saturates below $\sim$~1~K, with a value roughly five times greater than found for Tb$_2$Ti$_2$O$_7$\cite{99_gardner}. A single exponential muon spin relaxation from the sample is also observed at all temperatures. These findings are indicative of large, rapidly-fluctuating internal magnetic fields, as expected for a paramagnet\cite{97_dereotier}, and they are in good agreement with the decoupling scheme observed in the LF scan at 0.095~K. 
%CANNOT ESTIMATE Bi TO OBTAIN A VALUE FOR EFFECTIVE EXCHANGE     %CONSTANT. GARDNER USED VALUE FROM TB2MO2O7 IN HIS PAPER, WHICH  %SHOWS SPIN-FREEZING IN MUSR. I HAVE NO APPROPRIATE SAMPLE TO    %COMPARE TO. 
%SHOULD I ALSO DISCUSS WHAT FRACTION OF SPINS STOP IN THE SAMPLE %FOR THE MUSR MEASUREMENT? TYPICAL ASYMMETRY OF MUSR IN DR IS    %WHAT VALUE? 0.3?  

The absence of magnetic transitions in the bulk measurements, the presence of low-temperature short-range spin correlations from neutron diffraction, and the lack of any static magnetism detected by $\mu$SR down to 0.095~K provide strong evidence for a large moment spin liquid ground state in BaTb$_2$O$_4$. This exotic state likely arises from incredibly balanced $J_1$ and $J_2$ intrachain exchange interactions, which lead to a high degree of frustration. Single crystal inelastic neutron scattering and theoretical calculations are needed to help determine the crystal field level schemes and the evolution of the magnetic Hamiltonian throughout the AR$_2$O$_4$ series, as such work will lead to a deeper understanding of the origins of the exotic and varied magnetism in this family of materials.       

%_______________________________________________________________
%SHOULD EMPHASIZE THAT BATB2O4 IS A GOOD MODEL SYSTEM BECAUSE IT %SEEMS THAT ONLY VERY SHORT RANGE CORRELATIONS ARE NEEDED IN THE %MODEL

%HOWEVER, WITH REGARDS TO THE CLASSICAL CASE, SYSTEMS WITH LARGE %SPINS GENERALLY ORDER DUE TO WEAK INTERCHAIN INTERACTIONS OR   
%LARGE ANISOTROPY. NEXT TALK ABOUT HOW EXAMPLES OF SPIN LIQUIDS 
%IN CLASSICAL SYSTEMS ARE RARE, WITH A FEW PYROCHLORES BEING THE %MAIN CANDIDATES (SEE GARDNER REVIEW). SHOULD MENTION SOMETHING 
%ABOUT COMPARING CLASSICAL SPIN LIQUIDS TO THE QUANTUM VARIETY  
%FROM AN EXPERIMENTAL STANDPOINT AND HOW THAT IS USEFUL. ALSO   
%CAN MENTION HOW SUBLEADING TERMS IN THE HAMILTONIAN OFTEN      
%NEGLECTED BY THEORY LEAD TO CLASSICAL LRO (SEE DISCUSSION IN   
%2009 MENTRE PAPER AT THE END OF THE FIRST PARAGRAPH) AND       
%THEREFORE SPIN LIQUID CANDIDATES ARE HARD TO FIND.

%THEN TALK ABOUT (BA,SR)R2O4 SYSTEMS AND WHY THEY ARE GOOD      
%CANDIDATES FOR THIS PHYSICS AND OTHER NOVEL MAGNETIC GROUND    
%STATES. DESPITE THE 1D CHAINS IN THE SYSTEM, WEAK INTERCHAIN   
%COUPLINGS OR SINGLE ION ANISOTROPY TEND TO DRIVE THESE         
%MATERIALS TO LONG-RANGE ORDER AT LOW T. STILL, COMPLEXITIES    
%REMAIN (GO ON TO EXPLAIN THESE). 

%OTHER TWO LEG ZIGZAG LADDERS THAT I SHOULD LOOK INTO: CS2CUCL4, %KCUCL3, TLCUCL3, NH4CUCL3. THESE HAVE INEQUIVALENT RUNG        
%INTERACTIONS (TWO TYPES) AND ALL HAVE SPIN SINGLET GROUND      
%STATES.

\begin{acknowledgments}
This research was supported by the US Department of Energy (DOE), Office of Basic Energy Sciences. A.A.A. and V.O.G. were supported by the Scientific User Facilities Division. J.-Q.Y. and D.M. were supported by the Materials Science and Engineering Division. The neutron experiments were performed at the High Flux Isotope Reactor, which is sponsored by the Scientific User Facilities Division. Work at NHMFL-LANL was supported by NSF, US DOE and the State of Florida. Finally, we thank the staff of ISIS, where the $\mu$SR experiments were performed, for their hospitality.
\end{acknowledgments}

\end{document}